# An Observational Test of
# the Critical Earthquake Concept

D.D. BOWMAN[1], G. OUILLON[2,3], C.G. SAMMIS[1], A. SORNETTE[3] AND D. SORNETTE[2,3]

[1]*Department of Earth Sciences, University of Southern California, Los Angeles, CA 90089-0740*
[2]*Department of Earth and Space Sciences and Institute of Geophysics and Planetary Physics*
*University of California, Los Angeles, CA 90095-1567*
[3]*Laboratoire de Physique de la Matière Condensée*
*CNRS and Université de Nice-Sophia Antipolis, 06108 NICE Cedex 2, FRANCE*

**Abstract.** We test the concept that seismicity prior to a large earthquake can be understood in terms of the statistical physics of a critical phase transition. In this model, the cumulative seismic strain release increases as a power-law time-to-failure before the final event. Furthermore, the region of correlated seismicity predicted by this model is much greater than would be predicted from simple elasto-dynamic interactions. We present a systematic procedure to test for the accelerating seismicity predicted by the critical point model and to identify the region approaching criticality, based on a comparison between the observed cumulative energy (Benioff strain) release and the power-law behavior predicted by theory. This method is used to find the critical region before all earthquakes along the San Andreas system since 1950 with M ≥ 6.5. The statistical significance of our results is assessed by performing the same procedure on a large number of randomly generated synthetic catalogs. The null hypothesis, that the observed acceleration in all these earthquakes could result from spurious patterns generated by our procedure in purely random catalogs, is rejected with 99.5% confidence. An empirical relation between the logarithm of the critical region radius (R) and the magnitude of the final event (M) is found, such that log R ∝ 0.5 M, suggesting that the largest probable event in a given region scales with the size of the regional fault network.

## Introduction

Many investigators have attempted to use the methods of statistical physics to understand regional seismicity [*Rundle*, 1988a,b; *Rundle*, 1989a,b; *Smalley et al.*, 1985]. One approach has been to model the earthquake process as a critical phenomenon, culminating in a large event that is analogous to a kind of critical point [*Allègre and Le Mouel*, 1994; *Allègre et al.*, 1982; *Chelidze*, 1982; *Keilis-Borok*, 1990; *Saleur et al.*, 1996a; *Saleur et al.*, 1996b; *Sornette and Sornette*, 1990; *Sornette and Sammis*, 1995]. Attempts to link earthquakes and critical phenomena find support in the recent demonstration that rupture in heterogeneous media is a critical phenomenon [*Herrmann and Roux*, 1990; *Vanneste and Sornette*, 1992; *Sornette et al.*, 1992; *Lamaignere et al.*, 1996; *Andersen et al.*, 1997]. Also encouraging is the often reported observation of increased intermediate magnitude seismicity before large events [*Ellsworth et al.*, 1981; *Jones*, 1994; *Keilis-Borok et al.*, 1988; *Knopoff et al.*, 1996; *Lindh*, 1990; *Mogi*, 1969; *Raleigh et al.*, 1982; *Sykes and Jaumé*, 1990; *Tocher*, 1959]. Because these precursory events occur over an area much greater than is predicted for elasto-dynamic interactions, they are not considered to be classical foreshocks [*Jones and Molnar*, 1979]. While the observed long-range correlations in seismicity can not be explained by simple elasto-dynamic interactions, they can be understood by analogy to the statistical mechanics of a system approaching a critical point for which the correlation length is only limited by the size of the system [*Wilson*, 1979].

How does one test the critical point hypothesis for earthquakes? One of the fundamental predictions of this approach is that large earthquakes follow periods of accelerating seismicity. Previous investigators [*Bufe et al.*, 1994; *Bufe and Varnes*, 1993; *Saleur et al.*, 1996a; *Sammis et al.*, 1996; *Sornette and Sammis*, 1995; *Sykes and Jaumé*, 1990] have attempted to predict the time of the ensuing mainshock by quantifying the accelerating seismicity predicted by statistical physics. However, these works have all been case studies of individual events which may have been chosen precisely because they followed a period of accelerating seismicity. *Knopoff et al.* [1996] have conducted a systematic study that can be seen as a preliminary test of the critical point hypothesis. They found that all 11 earthquakes in California since 1941 with magnitudes greater than 6.8 were associated with anomalously high levels of intermediate ($M \geq 5.1$) earthquake activity. However, the five-year running- window method they used does not allow one to quantitatively test the existence of an acceleration in the seisimicity, if any. The spatial partition was made *a priori* between northern, central and southern California and no attempt was made to identify the regions which were approaching criticality. The purpose of the present paper is to test the existence of, and quantify the acceleration of, seismicity of intermediate earthquakes and to identify the critical region associated with each earthquake. Failure to observe a region of accelerating seismicity before these events would invalidate the critical point hypothesis.

It is important to note that the central prediction of the critical point hypothesis is that large earthquakes only occur when the system is in a critical state. The word "critical" describes a system at the boundary between order and disorder, and is characterized by both extreme susceptibility to external factors and strong correlation between different parts of the system. Examples of such systems are liquids and magnets, where the system progressively orders under small external changes. Critical behavior is fundamentally a cooperative phenomenon, resulting from the repeated interactions between "microscopic" elements which progressively "phase up" and construct a "macroscopic" self-similar state. Here, the important observation is the appearence of many different scales involved in the construction of the macroscopic rupture of a large earthquake.

The question now arises: how is this critical state related to the postulated self-organized criticality of the earth's crust [*Sornette and Sornette*, 1989;*Bak and Tang*, 1989; *Ito and Matsuzaki*, 1990; *Scholz*, 1991; *Sornette*, 1991; *Geller et al.*, 1997; *Main*, 1997]? *Huang et al.* [1998] have shown that these two concepts address different questions and describe properties of the crust at different time scales. Critical rupture occurs when the applied force reaches a critical value beyond which the system moves globally and abruptly, while self-organized criticality needs a slow driving velocity and describes the jerky steady-state of the system at large time scales. Thus, the critical rupture associated with large earthquakes constitutes a small fraction of the total history described by self-organized criticality. In their hierarchical model fault network, *Huang et al.* [1998] found that the critical nature of the large model-earthquakes emerges from the interplay between the long-range stress-stress correlations of the self-organized critical state and the hierarchical geometrical structure: a given level of the hierarchical rupture is like a critical point to all the lower levels, albeit with a finite size. The finite size effects are thus intrinsic to the process. Pushing this reasoning by taking into account the special role played by plate boundaries, *Sammis and Smith* [1997] have suggested that regional fault networks may not be in a state of continuous self-organized criticality. Rather, their analysis, performed on the same model as *Huang et al.* [1998], suggests that a large event may move the network away from criticality. A period of accelerating seismicity is then the signature of the approach back to the critical state. Furthermore, the large earthquake need not correspond exactly with the critical point. A large earthquake which is not immediately preceded by a period of accelerating seismicity may represent a region which had previously reached a state of self-organized criticality, but has not yet had an earthquake large enough to perturb the system away from the critical state. Alternatively, a region of accelerating seismicity which is not followed by a large earthquake is interpreted as a system which has achieved criticality, but in which a large event has not yet nucleated. The observation of accelerating seismicity is indirect evidence that the system has approached criticality and is therefore capable of producing a large earthquake. *Grasso and Sornette* [1997] have proposed another approach to test for criticality, which is to monitor the response of the crust to perturbations leading to induced seismicity. In this case, one tests directly the large "critical" susceptibility of the crust which should be present in the critical state.

For simplicity, in this paper it is implicitly assumed that large earthquakes occur at a time close to the critical point. We will test the critical point hypothesis by searching for a region of accelerating seismicity (i.e. a critical region) for all earthquakes stronger than magnitude 6.5 in California between the latitudes of 40°N (the northern extent of the San Andreas fault) and 32.5° N (the southern limit of the Southern California Seismic Network). The critical region is selected by optimizing the residue of the fit of the cumulative Benioff strain by a power law. We limit our study to events after 1950 in order to minimize artifacts in the data due to modifications of the seismic networks, such as the installation of the Southern California network in 1932. We also present detailed synthetic tests that have been performed in order to check the statistical significance of our results. While the acceleration pattern found for each single earthquake could be the result of chance with a probability close to 1/2, when taking all together, we find that our results reject the null hypothesis that the seismic acceleration is due to chance with a confidence level better than 99.5%. As a bonus, we find that the critical region size scales with the magnitude of the final event. Although the regional study defined above is complete for all events meeting the study criteria, the earthquakes cover a very limited magnitude range ($6.5 \le M \le 7.5$). In order to study the relationship between critical region size and the magnitude of the final event, we have extended the magnitude window by analyzing four additional events outside of the previously defined space/time/magnitude limits. These earthquakes, which include two large and two small events, help define the scaling region over a broader magnitude range ($4.7 \le M \le 8.6$). The proposed scaling law reinforces the critical point hypothesis, but will demand the analysis of many more cases to be put on a firm statistical basis.

**The approach to criticality**

*Bufe and Varnes*, [1993] and *Bufe et al.*, [1994] found that the clustering of intermediate events before a large shock produces a regional increase in cumulative Benioff strain, $\varepsilon(t)$, which can be fit by a power-law time-to-failure relation of the form

$$\varepsilon(t) = A + B(t_c - t)^m \qquad (1)$$

where $t_c$ is the time of the large event, B is negative and m is usually about 0.3. A is the value of $\varepsilon(t)$ when $t = t_c$, i.e. the final Benioff strain up to and including the largest event. The cumulative Benioff strain at time t is defined as

$$\varepsilon(t) = \sum_{i=1}^{N(t)} E_i(t)^{\frac{1}{2}}, \qquad (2)$$

where $E_i$ is the energy of the $i^{th}$ event and $N(t)$ is the number of events at time t. In calculating the energy release by an event, we follow the formulation of *Kanamori and Anderson* [1975], where

$$\log_{10} E = 4.8 + 1.5 M_s. \qquad (3)$$

*Bufe and Varnes*, [1993] justified (1) in terms of a simple damage mechanics model. However, *Sornette and Sammis*, [1995] pointed out that a power-law increase in the cumulative seismic strain release can also be expected for heterogeneous materials if the rupture process is analogous to a critical phase transition. In this approach, small and intermediate earthquakes are associated with the growing correlation length of the regional stress field prior to a large event. The final event in the cycle is viewed as being analogous to the critical point in a chemical or magnetic phase transition, occurring when the regional stress field is correlated at long wavelengths. Within this framework, the renormalization group method has been applied as a powerful tool for understanding the earthquake process in terms of the growing correlation length of the regional stress field [*Saleur et al., 1996a,b*].

The renormalization group method introduces the concept of a scaling region in the time and space before a large rupture. The essence of the renormalization group is to assume that the failure process at a small spatial scale length and temporally far from global failure (i.e. t « $t_c$) can be remapped (renormalized) to the process at a larger length scale and closer to the failure time. Thus, the damage mechanics model described above can be recognized as a mean-field approximation to the regional renormalization group (see *Saleur et al, 1996a,b* for a discussion of the application of the renormalization group to regional seismicity).

The regional clustering of intermediate events prior to a large earthquake can be understood in terms of the growing correlation length of the regional stress field. Here, "correlation length" refers to the capacity of the system to sustain a rupture. Early in the earthquake cycle (i.e., following an earlier large earthquake), the correlation length is short, reflecting a stress field which is very rough on the regional scale. Simply put, this implies that the distribution of stress on the regional scale is highly heterogeneous, and that the physical dimensions of regions capable of producing a rupture are smaller than the physical dimensions of the individual faults which are capable of producing a rupture. Thus, when an earthquake nucleates, it will reach the edge of the highly stressed patch before it can grow to become a large event. As these events occur, they smooth the stress field at long length scales by redistributing stress to neighboring regions, while roughening the stress field at short length scales through aftershocks. The process of smoothing the stress field at long wavelengths while roughening it at short wavelengths increases the correlation length of the regional stress field without shutting off small events. As the process continues, the growing correlation of the stress field allows ruptures in the smoothed stress field to grow to greater lengths, smoothing the stress field at even longer length scales. Thus, earthquakes which nucleate later in the cycle, and therefore in a more correlated stress field, are able to rupture barriers which would have halted the earthquake in an earlier, less correlated stress field. Only when criticality is reached is the stress field correlated on all scale lengths up to and including the largest possible event for the given fault network.

In addition to the accelerating seismic release in (1), the renormalization group approach makes observable predictions of the influence of local geology on the regional rate of seismic energy release. *Sornette and Sammis* [1995] pointed out that if the regional fault network is a discrete hierarchy, then the critical exponent m in (1) is complex, such that

$$\varepsilon(t) = Re[A + B(t_f\text{-}t)^{m+im'}] = A + B(t_f\text{-}t)^m\left[1 + C\cos\left(2\pi\,\frac{\log\,(t_f\text{-}t)}{\log\,\lambda} + \psi\right)\right] \qquad (4)$$

where A, B, and C are constants, $\lambda$ describes the wavelength of the oscillations in log-space, and $\psi$ is a phase shift which depends on the units of the time measurement. Note that the portion of the equation outside the square brackets is just a simple power-law. The expression within the square brackets in (4) is a correction due to the expansion of the imaginary component, and is a second-order effect which modulates the power-law.

Another implication of the critical point model is that small earthquakes are the agents by which longer stress correlations are established - they effectively smooth the stress field at larger scale lengths, as originally proposed by *Andrews*, [1975]. Also, the final catastrophic event in the sequence may occur in a system which does not have a single fault large enough to support the rupture. In such a scenario, the long range correlation of the stress field allows the final event to connect neighboring faults, as in the 1992 Landers earthquake. Alternatively, a network without a single largest element may be moved away from criticality by a cluster of intermediate magnitude events.

Finally, it is noteworthy that (1) contains no inherent spatial information. From a strictly theoretical viewpoint, the correlation length of the stress field could continue to grow up to the absolute physical size of the system. Thus it can be argued that this process is ultimately limited only by the size of the earth. A more physical limit is imposed by the size of the "local" fault network. For example, the physical extent of transform faulting along the Pacific-North American plate margin places a reasonable upper limit on the scale for which this process may occur in western North America. In a region with a more spatially extended network of faulting (e.g. China), the process could be expected to continue to even greater scales. Indeed, different domains within the plate margin may become critical independently of each other, each with their own scaling region in time and space. While these domains do have physical constraints on their size, they still are much larger than might be expected for actual triggering as an elastic process. The large length scales suggested by this process do *not* imply an actual triggering mechanism. Rather, they describe the evolution of a system which is sympathetic to increasingly large ruptures. The extreme length scales implied by this process make identification of the critical region difficult, and are one of the prime reasons why precursory increases in regional seismicity have not been widely recognized until recently.

**Spatial Identification of Critical Regions**

A fundamental prediction of the critical point model is that the seismicity in a region which is preparing for a large earthquake increases as power-law of the time to failure (Equation 1). Therefore, in the context of the critical point model, we can identify a region which is approaching criticality by optimizing the fit to (1). Initially, we will neglect the proposed log-periodic corrections to the power-law. Log-periodicity is a second-order phenomena related to the presence of a structural hierarchy, and is not a necessary condition for the definition of the critical region. However, we will qualitatively show that for most of the earthquakes we have considered in this study, log-periodicity is only observed in regions where the power-law is optimized.

We have developed a region optimization algorithm which identifies domains of accelerating seismicity. The seismicity in a test domain is fit to the power-law time-to-failure equation (1) and to a straight line corresponding to the null hypothesis that the seismicity rate is constant. Because the physics of criticality requires seismicity to be accelerating prior to the mainshock, we impose the conditions B<0 and 0<m<1 in (1). The critical exponent, m, is further constrained to be less than 0.8, because for $0.8 \leq m \leq 1$ the curvature of the power-law is virtually indistinguishable from that of a straight line over the time scales we consider. Finally, the date and magnitude of the final earthquake are assumed to be known quantities, which fixes the values of $t_f$ and A in (1) and allows a more accurate determination of the critical region.

In order to quantify the degree of acceleration in the seismicity, we define a curvature parameter C, where

$$C = \frac{\text{power-law fit root-mean-square error}}{\text{linear fit root-mean-square error}}. \qquad (5)$$

Therefore when the data is best characterized by a power-law curve, the RMS error for the power-law fit will be small compared to the RMS error of the linear fit, and C will be small. Conversely, if the seismicity is linearly increasing then the power-law fit will be statistically indistinguishable from a linear fit, and the parameter C will be at or near unity. Because the critical exponent is constrained to be less than 0.8, a region of linearly increasing seismicity will have a power-law fit whose RMS error is slightly greater than the error of the linear fit. This effect is greatest for regions of decelerating seismicity, which will have a power-law RMS much greater than that of the linear fit, corresponding to C much greater than unity.

The optimization procedure begins by isolating earthquakes within a small circle centered on the epicenter of the large earthquake under consideration. This smallest region is always chosen to include at least four events, so that the power-law can be uniquely determined. Within this region, the seismicity is fit to a straight line as well as the power-law, and the parameter C is calculated. The radius of the circle defining the current region of interest is then repeatedly increased by an arbitrary increment, and a new value of C is calculated at each step. After a predetermined number of steps, the curvature parameter C is plotted as a function of region size. The optimal radius of the critical region corresponds to the minimum value of C. An error estimate for the region determination can be found by measuring the width of the minimum in the plot of C.

For this study, we explore the region approaching criticality before historic earthquakes. We shall focus our study on all events in California with magnitudes greater than 6.5 between 32° N and 40°N latitude (Figure 1). Because the method is dependent on a complete catalog at small to intermediate magnitudes, we shall focus on events since 1950. The catalog used in this study was the Council of the National Seismic System (CNSS) Worldwide Earthquake Catalog, which is accessible via the World Wide Web at the Northern California Earthquake Data Center (http://quake.geo.berkeley.edu/cnss), and includes contributions from the member networks of the CNSS, including the Southern California Seismic Network, the Seismographic Station of the University of California at Berkeley, and the Northern California Seismic Network. To avoid ambiguities in catalogs at small magnitudes, the lower cutoff of the catalogs in this study is 2 units below the magnitude of the final event. In regions where the catalog is known to be complete to lower magnitudes, the cutoff is adjusted accordingly.

To illustrate our methodology, we now analyze the 1952 Kern County earthquake (Figure 2). Figure 3 shows the cumulative Benioff strain release for the three regions in Figure 2 and, for comparison, all of California. Figure 4 shows the corresponding plot of the optimum curvature parameter versus region size. The seismicity prior to the Kern County event was analyzed beginning in 1910. To account for changes in and variations between the seismic networks recording the catalogs, a lower magnitude threshold for seismicity prior to this event was imposed at M=5.5. At a radius of less than 200 km, there is not enough seismicity to obtain a statistically significant curve-fit. However, at a radius of 325 km, the seismicity produces a well-defined power-law. As the region expands, the pattern degrades. By 600 km the energy release is essentially a linear function of time. Thus, we conclude that the critical region is delimited by a circle with a radius of 325 km centered at 35°N, 119°W (Figure 2). The positive and negative errors on the region determination are defined as the radii where $c$ is greater than 25% of $1-c_{min}$. Note that the seismicity for all of California during the same time period (Figure 3d) shows no acceleration in activity; the power-law time-to-failure behavior is only observed within the critical region. Furthermore, notice that the cumulative seismicity has a structure which suggests log-periodic oscillation about the power-law in the 325 km radius region (Figure 3b), while the other regions show less organized fluctuations. For all of the earthquakes analyzed in this study, log-periodicity was only identifiable within the critical regions defined by the best-fitting power-law.

The procedure described above was repeated for all earthquakes greater than magnitude 6.5 since 1950 along the San Andreas system (Table 1a). Figure 5 and Figure 6 show the curvature versus region size for each earthquake and the corresponding cumulative Benioff strain release for each event's optimal region. There seems to be a weak correlation between the magnitude of an earthquake and the size of its associated critical region (Figure 7, circles). While there is significant scatter in the region sizes, large earthquakes tend to be preceded by larger critical regions than for small events. However, the relatively narrow range of magnitudes included in the study ($6.5 \leq M \leq 7.5$) make this observation rather tenuous.

To further explore the scaling of the critical region, we have analyzed a small set of earthquakes outside the space-time window of our study. These events span a wide range of magnitudes, and include the 1950 $M_w$=8.6 Assam earthquake [*Triep and Sykes*, 1997], the 1906 $M_w$=7.9 San Francisco earthquake [*Sykes and Jaumé*, 1990] 1986 $M_L$=5.6 Palm Springs earthquake, and the 1980 $M_b$=4.8 Virgin Islands earthquake [*Varnes and Bufe*, 1996].

Notice that the Assam, San Francisco and Virgin Islands earthquakes each correspond probably to the upper levels of the corresponding regional hierarchy of earthquakes. This is the most favorable situation to clearly qualify a critical behavior as shown in *Huang et al.*, [1998]. For smaller earthquakes down the hierarchy, the relative size of the fluctuations of the Benioff strain is larger due to the influence of the (preparatory stages of) earthquakes at the upper levels of the hierarchy. Critical behavior is also expected to describe the maturation of intermediate earthquakes, such as the Palm Springs earthquake, however with increasing uncertainty due to the shadow cast by the larger events. This is our justification for not presenting a systematic study of the many intermediate earthquakes, which is left for a future work. Each of these events were analyzed with the above procedure using the CNSS Catalog (Figure 8, Table 1b). These earthquakes, which cover a much larger magnitude range ($4.8 \leq M \leq 8.6$), show a very well defined trend for large events to correspond with large region sizes (Figure 7).

**Statistical tests on synthetic catalogs**

By its very definition, our procedure will identify a specific pattern (the power law acceleration of seismicity measured by the Benioff strain) in a complex spatio-temporal set using an optimization procedure. The question arises whether this optimization would pick up a power-law pattern in any random set of earthquakes. In other words, a power-law acceleration might always be found provided that we give sufficient flexibility (here in the size R of the region) in the selection of the subset of the random set. Loosely speaking, Ramsey's theorem states that it is always possible to find any pattern we want in a sufficiently large random set [*Graham and Spencer*, 1990; *Graham et al.*, 1990]. For instance, this theorem explains why the ancients found recognizable everyday life structures in star constellations. Could it thus be that the observed acceleration in all of these earthquakes results from spurious patterns generated by our procedure?

In order to test this null hypothesis, we have generated 1000 synthetic catalogs, each containing 100 earthquakes. Each catalog corresponds to a given seismic history culminating in a main shock. The (intermediate) earthquakes are placed at random and uniformly within a square [-1000;1000] by [-1000;1000]. Each event occurs at a random time uniformly taken in the interval [0, 1000], and is characterized by a magnitude taken in the interval [5.5; 7.5] according to the Gutenberg-Richter law (log N = a - b M) with b=1. Finally, the main shock is positioned at the center (0,0) of the square in space and at the end of the time window $t_c$=1000. We have considered two magnitudes for the main shock (7.5 and 8.5) in order to test the sensitivity of our results with respect to this parameter. We use the above optimization procedure and construct the cumulative Benioff strain which is fit to equation (1), with A fixed, $t_c$ fixed and the exponent m constrained between 0 and 1.

Figure 9 shows the cumulative number N(C) of synthetic catalogs whose best curvature parameter over all possible radii R is smaller than or equal to C (N(C) is thus a conditional probability). The inset shows the full range of C-values while the main figure focuses on the most interesting values of C smaller than 1, which qualify that the power law (1) provides a better fit and therefore signals an acceleration. The two curves correspond to the two choices for the magnitude of the main shock and show the sensitivity of the results with respect to fixing the parameter A in equation (1) in the selection of the region. To compare with the genuine catalogs, we choose the case where the optimal C is always found smaller than or equal to 0.7. The probability that a random catalog gives an optimal C smaller than 0.7 is seen in Figure 9 to be slightly smaller than 0.5 for the main shock magnitude equal to 7.5 and smaller than 0.4 for the main shock magnitude equal to 8.5. This result confirms our suspicion that the application of an optimization procedure on a single catalog has close to 1/2 probability to generate a spurious acceleration out of pure randomness. However, the joint probability that eight earthquakes, chosen arbitrarily (i.e. without picking them *a priori*), all exhibit an optimal C less than or equal to 0.7 is less than $(1/2)^8 \leq 0.004$. This shows that the null hypothesis that the observed acceleration in all these eight earthquakes could result from spurious patterns generated by our procedure in purely random catalogs is rejected at a confidence better than 99%.

Finally, we note that the procedure proposed here and our statistical test are designed solely to test for periods of accelerating seismicity before large earthquakes. The results of this study show that the space-time distribution of seismicity is not purely random, as has been implied by models of self-organized criticality (e.g. *Bak and Tang* [1989], *Main* [1997], and *Geller et al.* [1997]). Rather, we have observed a systematic, non-spurious deviation from random behavior. While this observation does not constitute proof of the critical earthquake hypothesis, it is consistent with the conceptual model originally proposed by *Sornette and Sammis* [1995].

**Discussion**

It is important to examine the influence of the constraints that we have used in the fits. Fixing A as given by the main shock is justified a priori by the better quality of the fit and the fewer number of fitting parameters. However, in principle, one would like to free A and thus be able to estimate the magnitude of the main shock (which is directly related to A) from the accelerated seismicity. We have thus tested the influence of removing the constraint on A and on the exponent m. For Kern County, we find the smallest curvature parameter for a larger optimal radius 800±100 km instead of 325 ± 75 km (see table 1a) with an exponent around 1.5 (thus larger than 1). This reflects a decelerating trend instead of an acceleration, a result which holds when taking into account the intermediate earthquakes above magnitude 5.5. If we move the lower threshold for the selection of precursors to magnitude 6, we recover the results of table 1a with the same optimal radius and the same exponent. This shows the sensitivity of the three parameter fit with such sparse data. Similarly, for the Landers earthquake we find an optimal radius of 300±50 km  instead of 150±15 km (table 1a), again with an exponent close to 1.5.   However, in this case there are not enough events to increase the threshold magnitude as was done for Kern County.

We conclude that, at this stage of exploratory research, the sparseness of the catalogs prevents the realization of the full program which would consists in "predicting" completely the main shock properties (magnitude from A, time of occurrence from $t_c$ and position by varying the position of the circle). This would lead to fits with too many parameters to give robust results. The procedure developed above is thus justified by the minimization of the number of fitting parameters to test in the most robust possible way the critical earthquake hypothesis.  Some previous works [*Sornette and Sammis*, 1995; *Bufe and Varnes*, 1996] have been more ambitious and have fit the cumulative Benioff strain to the log-periodic equation (4), which contains even more fitting parameters. But this was done only on a few case studies and no systematics were obtained. It is possible that equation (4) might provide better results than are reported here because, notwithstanding the larger number of parameters, the log-periodic structures provide stringent constraints on the fits. This is currently being investigated and will be reported elsewhere.

We now turn to a possible interpretation of figure 7, which suggests that the logarithm of the critical region radius scales directly with the magnitude of the final event in the sequence.  A line with a slope of $\frac{1}{2}$ gives an excellent fit to the data. *Dobrovolsky et al.* [1979] and *Keilis-Borok and Kossobokov* [1990] report a similar scaling log R = 0.43 M for the maximum distance between an earthquake and its precursors, based on a completely different procedure, namely the optimization of pattern recognition techniques [*Gelfand et al*, 1976]. However, is there a theoretical justification for such a relationship?  One simple possibility is that the energy of the final event scales with the volume of the crust approaching criticality, such that

$$R^3 \propto E \tag{6}$$

where E is the energy of the final event and R is the radius of the critical region. *Kanamori and Anderson* [1975] showed that log $E \propto \frac{3}{2} M_s$, so that

$$\log R^3 = 3 \log R \propto \log E \propto \frac{3}{2} M, \tag{7}$$

which yields

$$\log R \propto \frac{1}{2} M. \tag{8}$$

However, an implicit assumption of this derivation is that the radius of the critical region is equal to or less than the thickness of the seismogenic crust. While this may be an acceptable assumption for small earthquakes, events larger than approximately magnitude 6 rupture the entire crust. For these events, the radius of the critical region is much larger than the thickness of the seismogenic zone, and the process must be treated as approximately 2 dimensional. Thus, for large events the scaling in (6) should be $R^2 \propto E$, which yields a slope of $\frac{3}{4}$, much higher than the observed slope of $\frac{1}{2}$.

The critical point hypothesis suggests another explanation for the slope of $\frac{1}{2}$ observed in Figure 7. In this model, the correlation length of the regional stress field is not constrained by elastic interactions, but rather by the physical dimensions of the system. Thus, it might be expected that the critical region is controlled by the size of the regional fault network, rather than by the transfer of elastic energy to the fault plane. Assuming that fault networks are self-similar [*Hirata*, 1989; *King*, 1986; *King et al.*, 1988; *Ouillon et al.*, 1996; *Robertson et al.*, 1995; *Turcotte*, 1986], then the linear dimension of the fault network should scale with the length of its largest member (L), i.e. $R \propto L$. This relation of proportionality does not mean that R and L are similar in order of magnitude, but only that R is proportional to L. In practice, R can be ten times larger than L or more. If the largest event occurs on the largest coherent element of the network, then $M_s \propto 2$ log L [*Kanamori and Anderson*, 1975], so that

$$2 \log R \propto 2 \log L \propto M \qquad (9)$$

and

$$\log R \propto \frac{1}{2} M. \qquad (10)$$

While the critical point model implies a slope of $\frac{1}{2}$ in Figure 7, a least-squares fit to the region size-magnitude plot gives a slope of 0.44. There is also considerable scatter about this best-fit line, particularly among the intermediate magnitude events (M ≈ 6.5). This may be in part due to the simplified data analysis procedures that were used in this study and also to the expected cross-over between 2D to 3D expected around this magnitude. Our study indicates that the 2D-3D cross-over is not only dictated by the comparison of the main shock rupture size to the seismogenic thickness. Our finding suggests another important scale, namely the size R of the critical region in which the maturation of the main shock occurs. Since R is significantly larger than L, the cross-over should be very smooth and extend over a large magnitude range. This seems to be confirmed by a recent careful reanalysis of the Gutenberg-Richter law for Southern California which found no cross-over up to the largest magnitude 7.5 [*Sornette et al.*, 1996].

The present work can be improved in many ways. First, there is no physical reason for the critical region to be a perfect circle centered on the epicenter of the final event. We have already extended this analysis to use elliptical regions rather than circles, with no significant modification of the results. A better approach would be to use the natural clustering of the seismicity to define the regions. The seismicity could also be much more accurately characterized by incorporating the second-order log-periodic corrections in (4). These fluctuations permit a more precise characterization of the precursory seismicity, which would consequently enable a more accurate determination of the critical region. Such procedures are clearly more computationally intensive, and are reserved for a later study. Finally, this study has made the explicit assumption that the final earthquake occurs as soon as the system achieves criticality. Some of the uncertainty in fitting equation (1) could be removed by incorporating an arbitrary time delay into the equation describing the accelerating seismicity.

In spite of these simplifications, it is encouraging that all of the events on the San Andreas fault system greater than magnitude 6.5 in the last half century were preceded by an identifiable region of accelerating seismicity which follows a power-law time-to-failure. Furthermore, the derived region size seems to follow a physically reasonable scaling with the magnitude of the final event which can be understood within the framework of the critical point model.

## Acknowledgements

We would like to acknowledge the helpful comments by the reviewers Peter Mora and Jim Brune, as well as the Associate Editor, Ruth Harris. We would also like to thank Bernard Minster and Nadya Williams for their comments on the manuscript. This work was funded by NSF grant #EAR-9508040 (DDB), partially supported by NSF EAR9615357 (DS) and by the Southern California Earthquake Center (CGS). SCEC is funded by NSF Cooperative Agreement EAR-8920135 and USGS Cooperative Agreement 14-08-0001-A0899 and 1434-HQ-97AG01718. This paper is SCEC contribution # 417 and IGPP contribution #XXXX.

**Figure Captions**

**Figure 1.** All earthquakes in California since 1950 with M ≥ 6.5 and with epicenters between 32° N and 40°N latitude. (1) 1952 $M_w$=7.5 Kern County. (2) 1968 $M_w$=6.5 Borrego Mountain. (3) 1971 $M_w$=6.6 San Fernando. (4) 1983 $M_w$=6.7 Coalinga. (5) 1987 $M_w$=6.6 Superstition Hills. (6) 1989 $M_w$=7.0 Loma Prieta. (7) 1992 $M_w$=7.3 Landers. (8) 1994 $M_w$=6.7 Northridge.

**Figure 2.** Three examples of the regions tested for accelerating seismicity before the 1952 Kern County earthquake. The optimal critical region is the dark circle at a radius of 325 km.

**Figure 3.** Cumulative Benioff strain (Equation 2) before the 1952 Kern County earthquake. (a) Seismicity within a 200 km radius. The power-law and linear fits both fit the data equally well. (b) Seismicity within a 325 km radius. The power-law fits the data significantly better than a line. Note the suggestion of log-periodic fluctuations about the power-law. (c) Seismicity within 600 km radius. Neither the power-law nor the linear fit adequately describe the seismicity. (d) Seismicity for all of California. The seismicity is linearly increasing.

**Figure 4.** Curvature paramter C as a function of region radius for the Kern County earthquake. Smaller values of C reflect a better power-law fit. The range of radii with the best power-law fit are shaded. The minimum is at a radius of 325 km.

**Figure 5.** Curvature paramter C as a function of region radius for all of the events meeting the study criteria (see Figure 1).

**Figure 6.** Cumulative Benioff strain release for the optimum regions in Figure 5.

**Figure 7.** Optimal region radius as a function of the magnitude of the final event. The best fitting line has a slope of 0.44.

**Figure 8.** Curvature parameter C as a function of region radius and cumulative Benioff strain for the optimal region for (a) 1950 $M_w$=8.6 Assam, (b) 1906 $M_w$=7.9 San Francisco, (c) 1986 $M_L$=5.6 North Palm Springs, and (d) 1980 $M_L$=4.8 Virgin Islands earthquakes.

**Figure 9:** Cumulative number N(C) of synthetic catalogs whose best curvature parameter over all possible radii R is smaller than or equal to C. The inset shows the full range of C-values while the main figure focuses on the most interesting values of C smaller than 1, which qualify that the power law (1) provides a better fit and therefore signals an acceleration. Circles: Main shock magnitude is 7.5; Squares: main shock magnitude is 8.5.

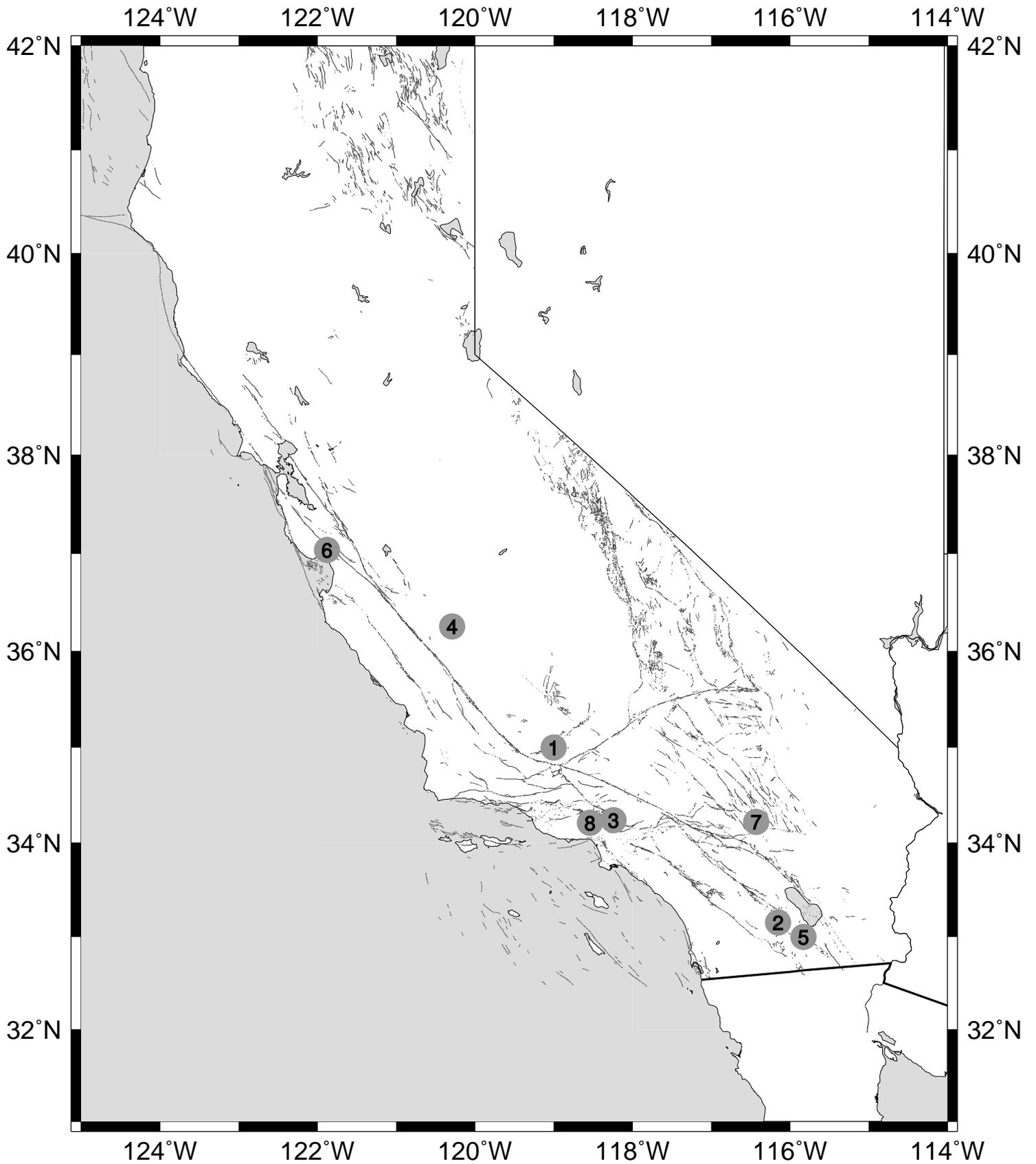

## Table 1a

| Earthquake | Date | Magnitude | Region Radius (km) | m |
|---|---|---|---|---|
| Kern County | 7/21/52 | 7.5 | 325±75 | 0.3 |
| Landers | 6/28/92 | 7.3 | 150 ±15 | 0.18 |
| Loma Prieta | 10/18/89 | 7.0 | 200±30 | 0.28 |
| Coalinga | 5/2/83 | 6.7 | 175±10 | 0.18 |
| Northridge | 1/17/94 | 6.7 | 73±17 | 0.1 |
| San Fernando | 2/9/71 | 6.6 | 100±20 | 0.13 |
| Superstition Hills | 11/24/87 | 6.6 | 275±95 | 0.43 |
| Borrego Mountain | 4/8/68 | 6.5 | 240±60 | 0.55 |

## Table 1b

| Assam | 8/15/50 | 8.6 | 900±175 | 0.22 |
|---|---|---|---|---|
| San Francisco | 4/18/06 | 7.7 | 575±240 | 0.49 |
| Palm Springs | 7/8/86 | 5.6 | 40±5 | 0.12 |
| Virgin Islands | 2/14/80 | 4.8 | 24±2 | 0.11 |

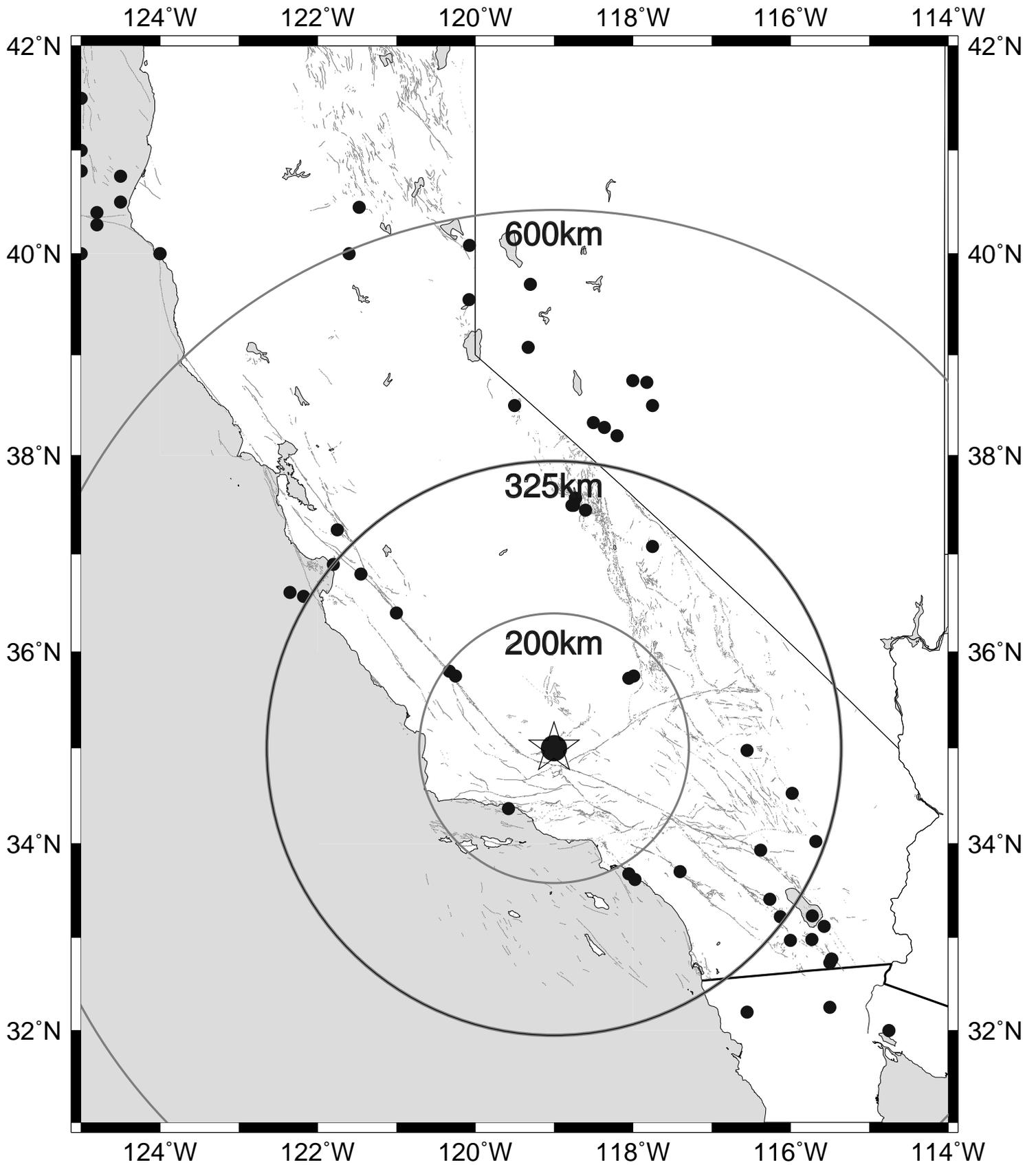

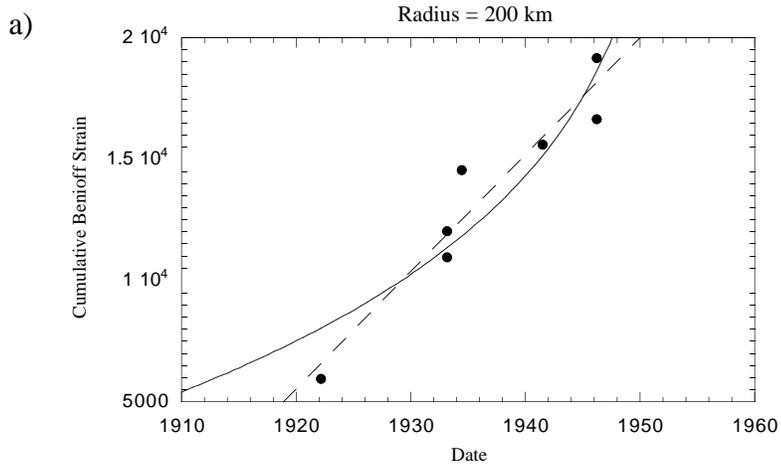

a)

Radius = 200 km

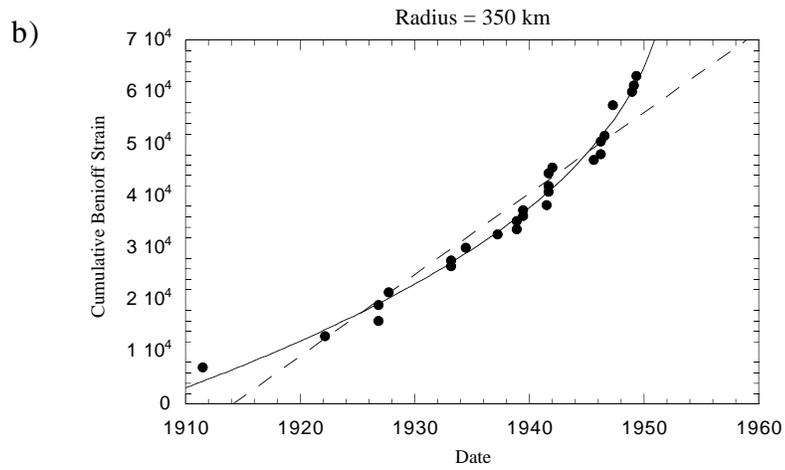

b)

Radius = 350 km

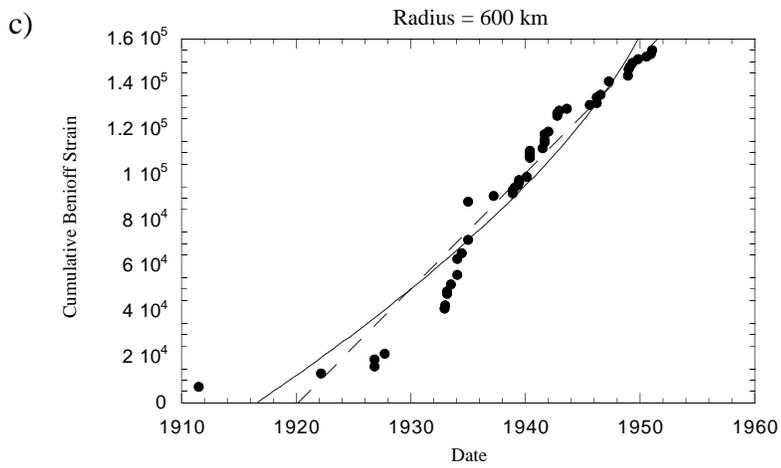

c)

Radius = 600 km

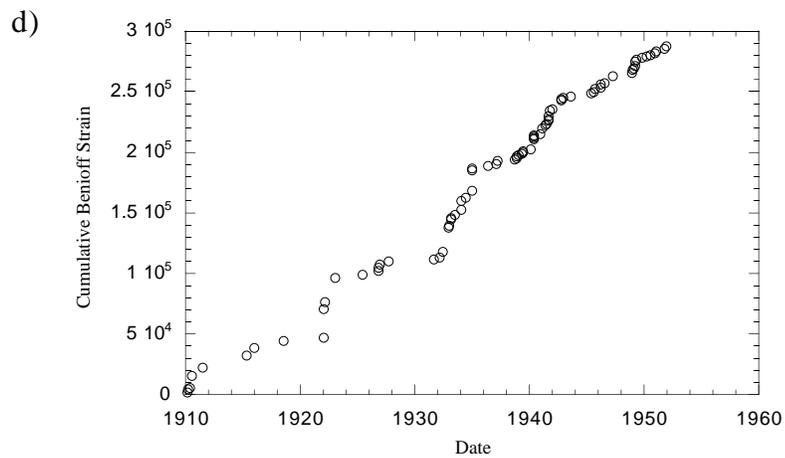

d)

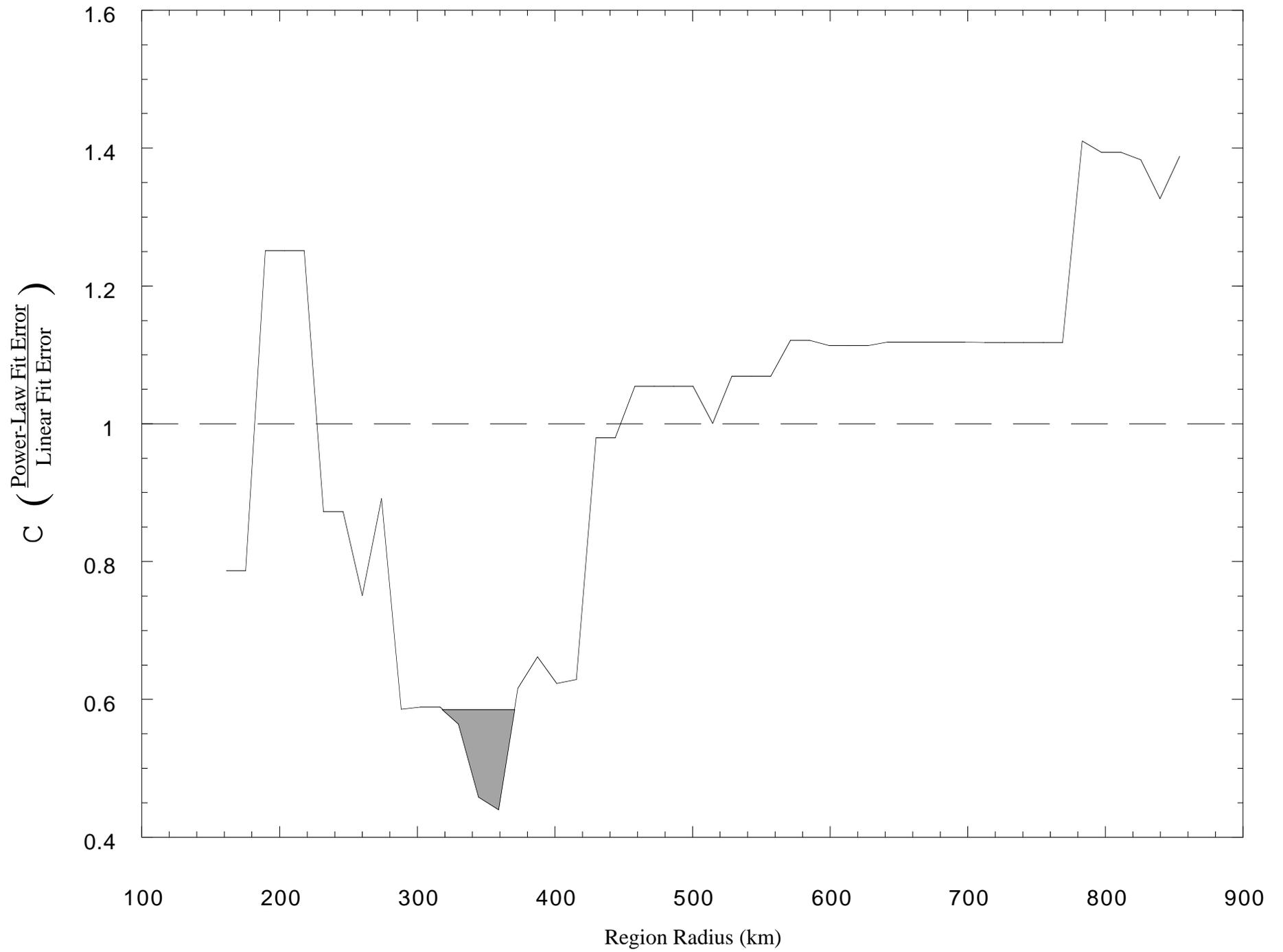

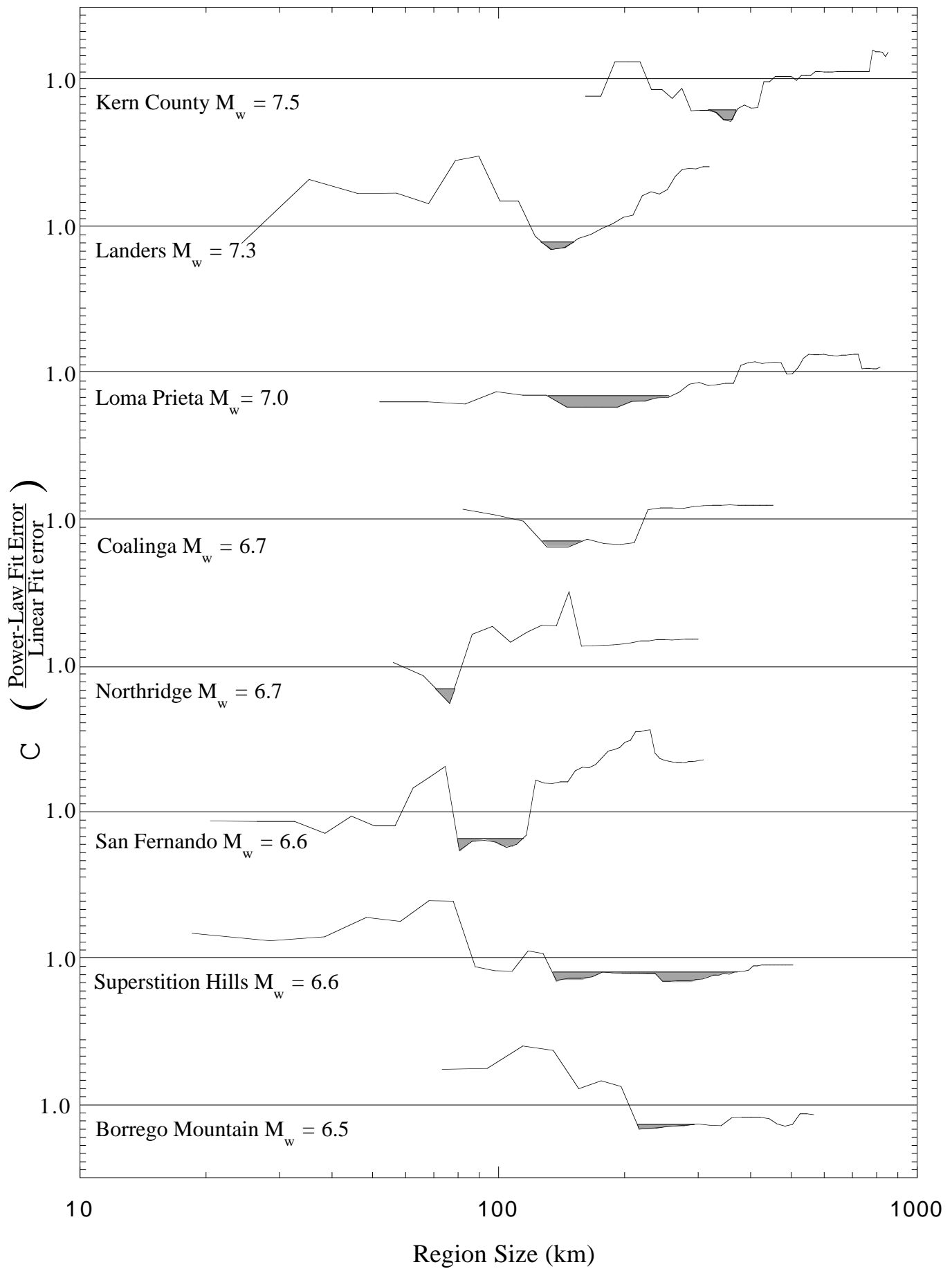

Kern County $M_w = 7.5$

Landers $M_w = 7.3$

Loma Prieta $M_w = 7.0$

Coalinga $M_w = 6.7$

Northridge $M_w = 6.7$

San Fernando $M_w = 6.6$

Superstition Hills $M_w = 6.6$

Borrego Mountain $M_w = 6.5$

$C \left( \dfrac{\text{Power-Law Fit Error}}{\text{Linear Fit error}} \right)$

Region Size (km)

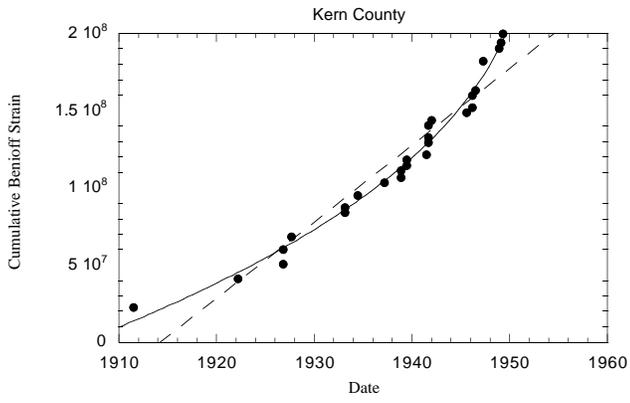

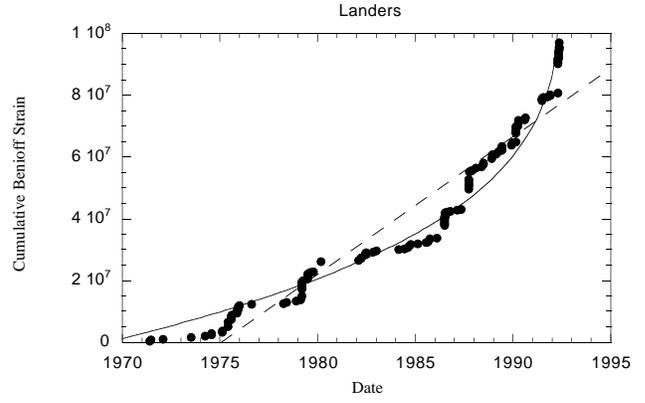

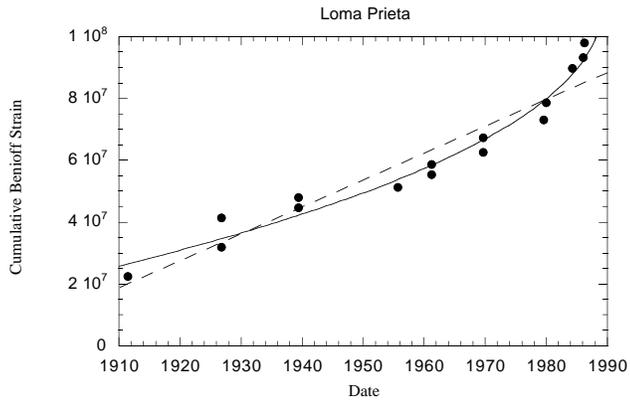

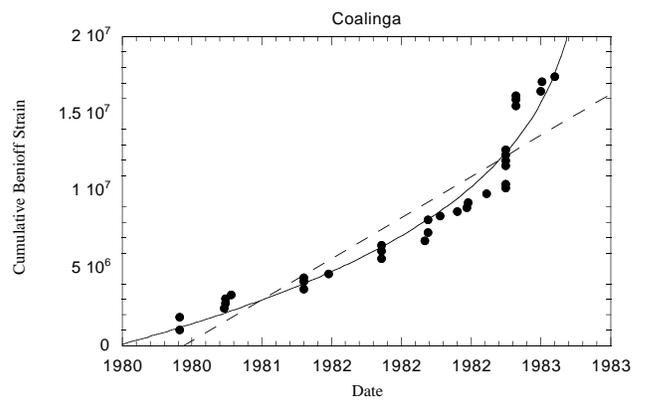

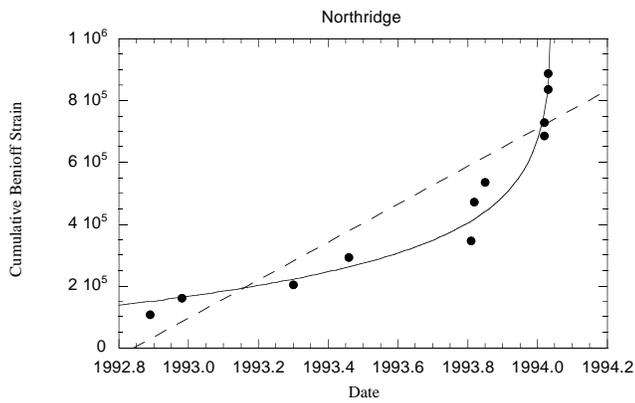

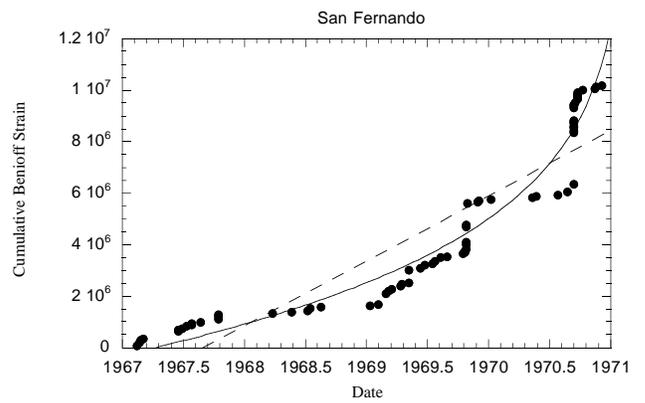

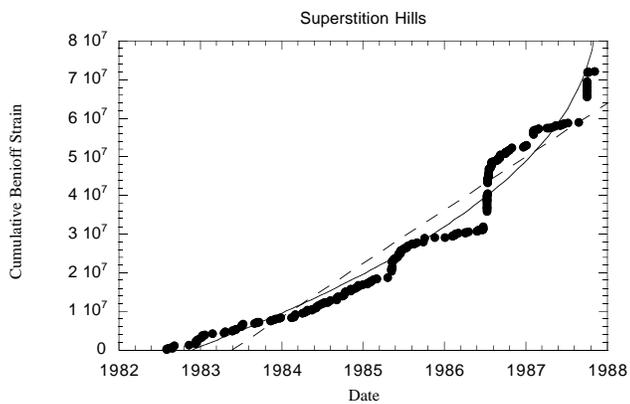

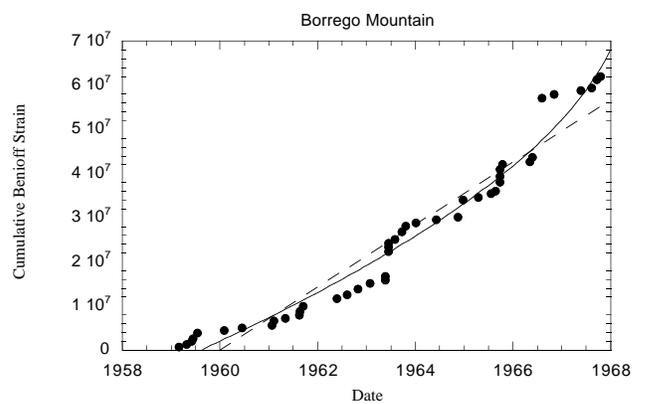

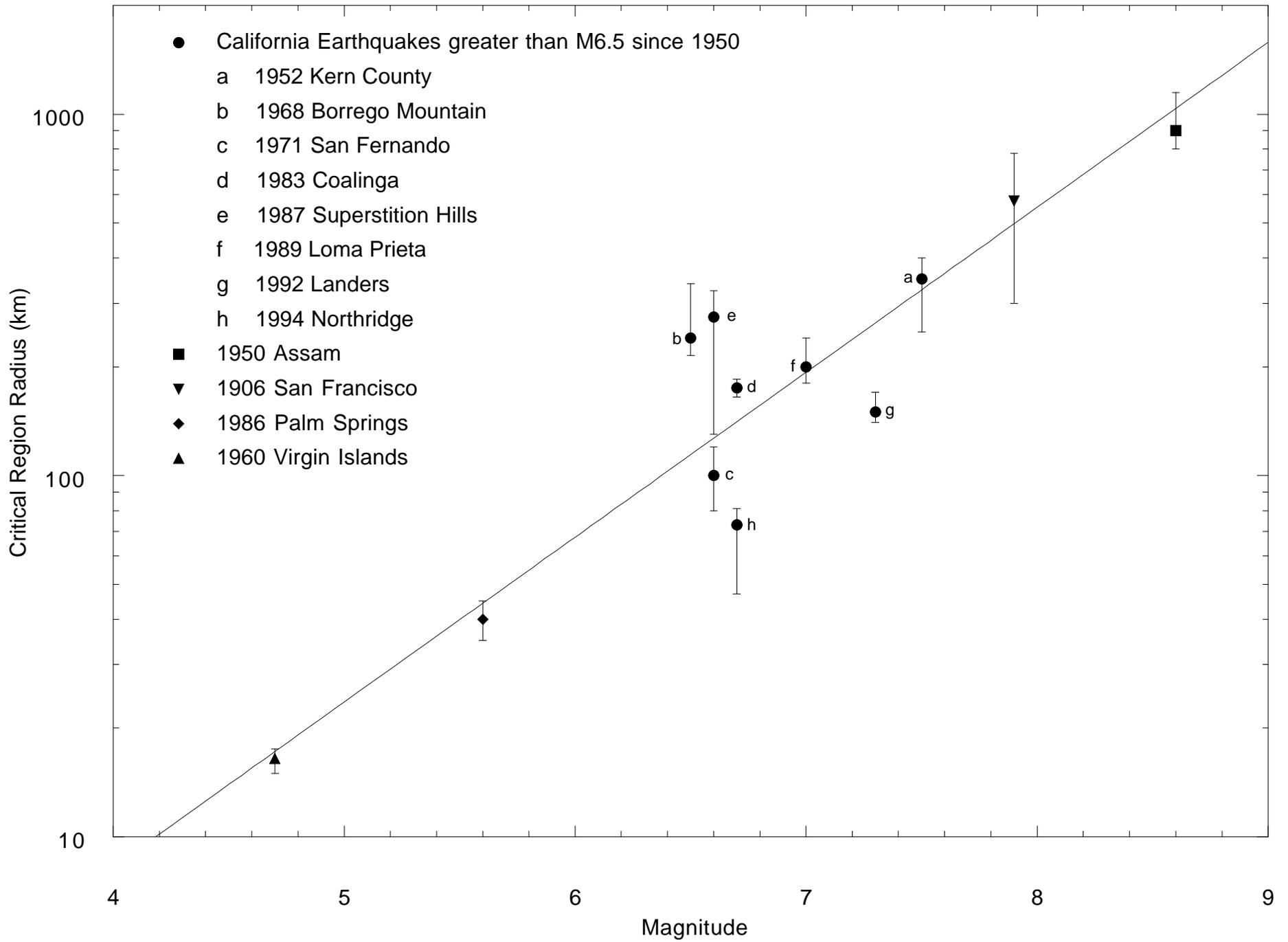

## a) Assam

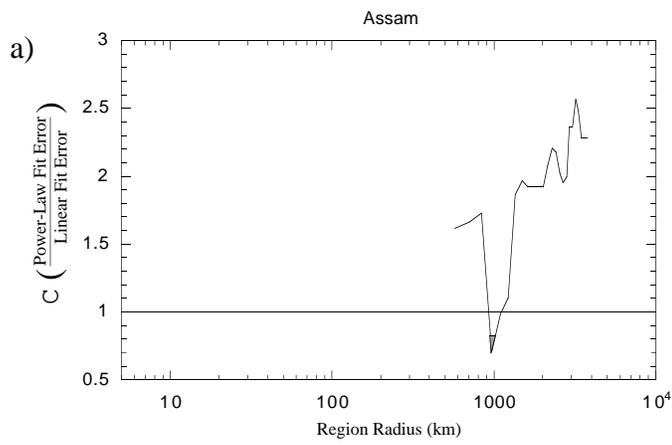
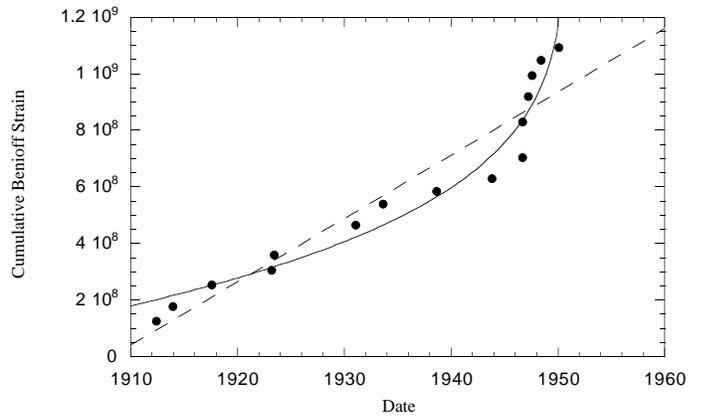

## b) San Francisco

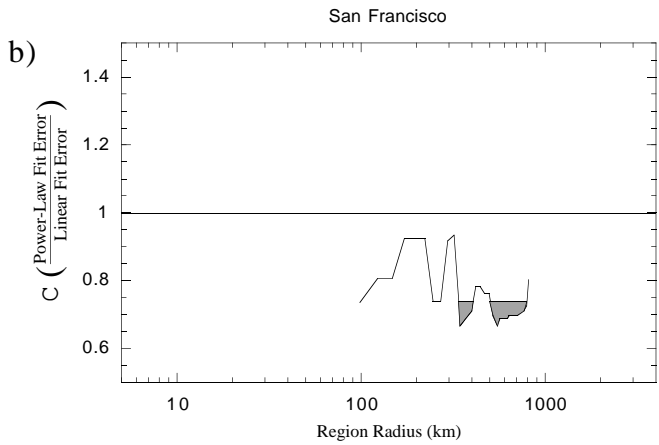
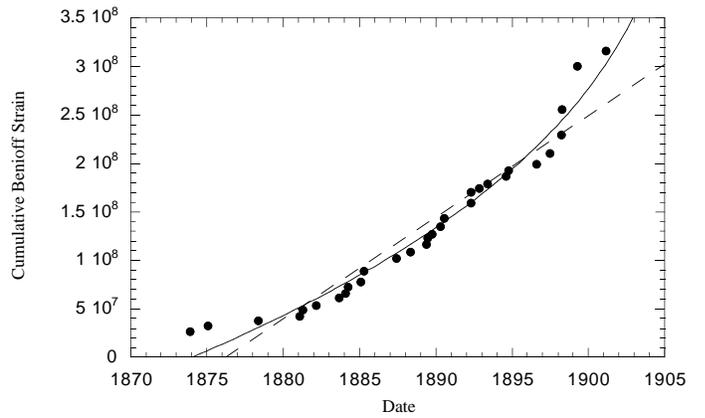

## c) Palm Springs

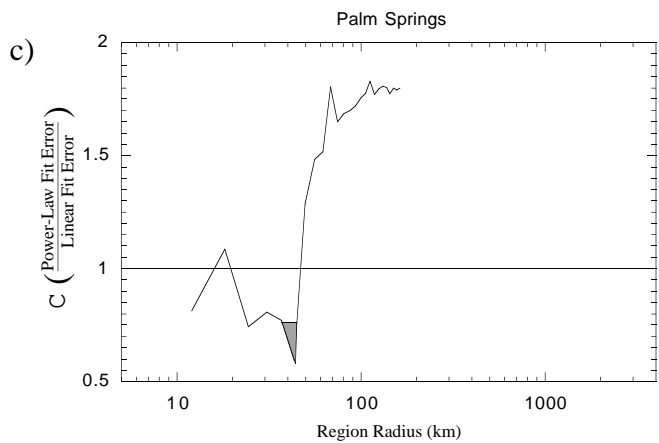
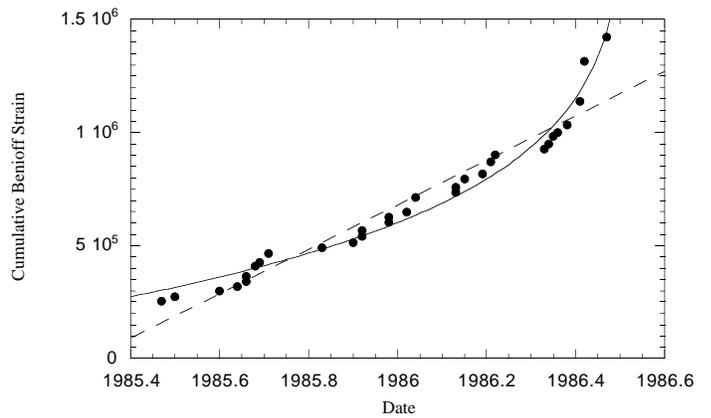

## d) Virgin Islands

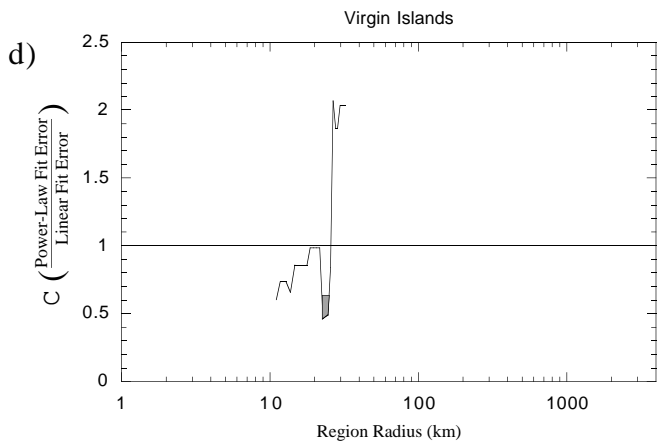
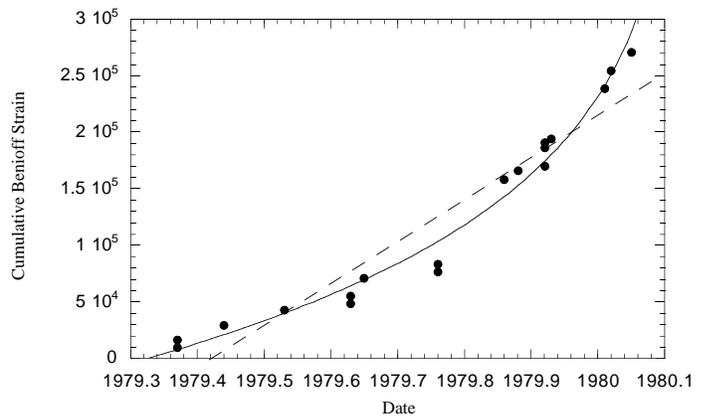

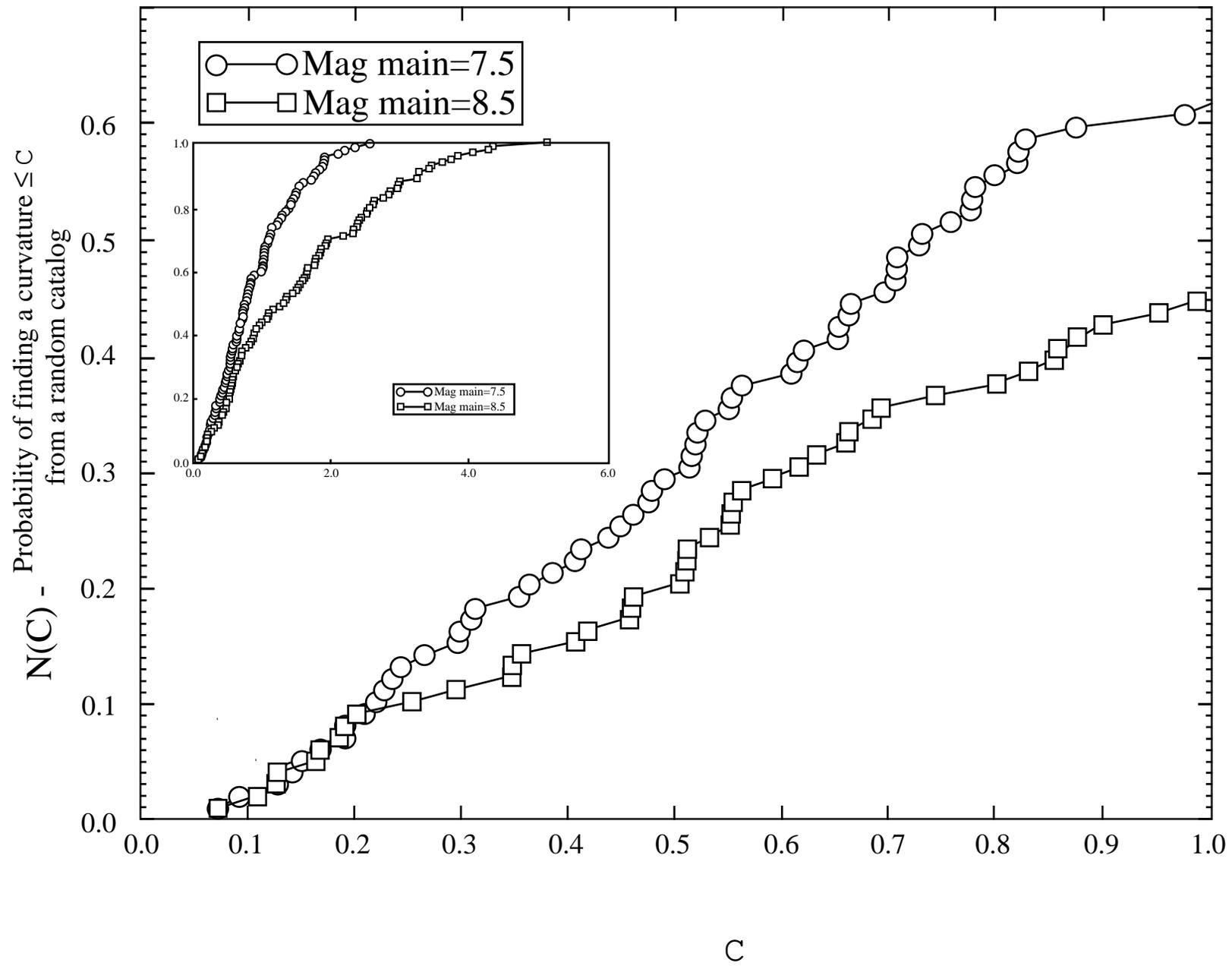